\documentclass[
 reprint,
 amsmath,amssymb,
 aps,
 prb,
]{revtex4-2}

\usepackage{graphicx}
\usepackage{dcolumn}
\usepackage{bm}
\usepackage{xcolor}

\begin{document}

\title{\textbf{Analytical Theory of Photon Tunneling and Near-Field Heat Transfer \\ Between Dissimilar Materials}}

\author{Kartika N. Nimje}
\affiliation{ICFO -- Institut de Ci\`encies Fot\`oniques, The Barcelona Institute of Science and Technology, Mediterranean Technology Park, Av. Carl Friedrich Gauss 3, Castelldefels, 08860, Barcelona, Spain}

\author{Mariano Pascale}
\affiliation{ICFO -- Institut de Ci\`encies Fot\`oniques, The Barcelona Institute of Science and Technology, Mediterranean Technology Park, Av. Carl Friedrich Gauss 3, Castelldefels, 08860, Barcelona, Spain}

\author{Georgia T. Papadakis}
\email{georgia.papadakis@icfo.eu}
\affiliation{ICFO -- Institut de Ci\`encies Fot\`oniques, The Barcelona Institute of Science and Technology, Mediterranean Technology Park, Av. Carl Friedrich Gauss 3, Castelldefels, 08860, Barcelona, Spain}

\date{\today}

\begin{abstract}
Near-field radiative heat transfer can exceed the blackbody limit through evanescent-mode coupling across nanoscale gaps. This enhancement underpins applications including thermophotovoltaic energy conversion, electroluminescent cooling, thermal rectification, and photon absorption in plasmon-assisted photodetection. These systems most often involve photon- or heat-exchange between dissimilar interfaces, particularly between a semiconductor and a metal. Despite the prevalence of this asymmetric configuration, no closed-form description of its near-field interaction exists. Here, we derive a closed-form analytical description of photon tunneling that clarifies the roles of material properties, namely the plasma frequency, optical loss, and semiconductor absorption, in the thermal exchange. We show that the dominant in-plane wave vector of the radiative heat transfer is an approximate average of the corresponding values for two symmetric reference systems: a plasmonic--plasmonic cavity and a semiconductor--semiconductor cavity. These results establish a compact analytical framework for near-field heat transfer between dissimilar materials.

\end{abstract}

\maketitle

\section{Introduction}

When two bodies are placed in close proximity with respect to the characteristic thermal wavelength, the way they exchange radiative heat changes fundamentally with respect to the blackbody limit. In particular, evanescent waves and surface-polariton modes can tunnel across a nanometric vacuum gap separating two surfaces, yielding heat transfer rates that can exceed the blackbody prediction~\cite{polder_theory_1971,joulain_surface_2005}. This has motivated efforts to understand and control the spectral and modal content of near-field heat transfer~\cite{biehs_near-field_2021}. However, analytical descriptions are challenging to derive, as computing radiative heat generally requires complex integration over spatial and spectral frequencies of photonic interactions. Nonetheless, for the \textit{symmetric} problem of radiative heat exchange between \textit{identical} surfaces, asymptotic and closed-form expressions have been derived under suitable approximations~\cite{rousseau_asymptotic_2012,iizuka_analytical_2015,pascale_tight_2023}. These serve to understand the role of coupled surface modes, optical loss, and material dispersion in near-field heat transfer. In particular, the original Polder--van Hove formalism provides a mode-resolved description of photon tunneling between planar bodies~\cite{polder_theory_1971}, whereas subsequent analytical work related the heat flux to coupled surface-mode dispersion~\cite{iizuka_analytical_2015}. More recently, bounds on near-field radiative transfer clarified how material response constrains the maximum thermal exchange~\cite{benabdallah_fundamental_2010}.

With respect to practical applications, near-field radiative heat transfer promises improved performance in nanoscale thermal-management devices, such as near-field energy-conversion platforms, e.g. thermophotovoltaics, electroluminescent cooling, photonic thermal rectification, sensing, and thermal storage. For the majority of these applications, radiative heat exchange occurs between two \textit{dissimilar} materials~\cite{otey_thermal_2010,tang_near-field_2020,mittapally_near-field_2021}, in other words within an \textit{asymmetric} cavity. In these practical cases, large overall heat transfer is not always useful. By contrast, selectivity over spectral range, bandwidth, dominant in-plane wave vector, and modal distribution is often a key requirement. For instance, in the case of near-field thermophotovoltaics (TPVs) consisting of a low-bandgap PV cell in the vicinity of a hot metallic emitter~\cite{zhao_high-performance_2017,mittapally_near-field_2021}, sub-bandgap photons parasitically heat the cell, while photons far above the bandgap increase thermalization losses. The relevant design objective is therefore not simply to maximize heat transfer, but to spectrally align the dominant tunneling contribution with the photovoltaic band edge. Similar requirements apply to the other aforementioned applications, the majority of which involve a semiconducting interface, serving as either light-emitting diode (LED) or PV cell, and a metallic one (often plasmonic). 

Despite the relevance of the asymmetric PV/LED-metal configuration, there exists no closed-form solution to the radiative heat exchanged between a semiconductor and a plasmonic metal in the near-field, which optically emulates it. However, some prior work exists in special cases; for instance, a reduced analytical model was developed in the case of near-field TPV systems, demonstrating that appropriate light management can optimize, simultaneously, the extracted electrical power density, the open-circuit voltage, and the conversion efficiency~\cite{papadakis_thermodynamics_2021}. Furthermore, coupled-mode approaches serve to design resonant near-field TPV systems while accounting for geometry-induced modes in structured emitters~\cite{wang_near-field_2017_cmt}. More recently, the photon-tunneling mechanism has been analyzed within the context of near-field TPV systems with a plasmonic emitter, in terms of total-internal-reflection and self-coupled surface-plasmon-polariton modes~\cite{li_photon_2025}. Similar results have been derived for near-field electroluminescent cooling, where the thermal exchange occurs between an LED and a photovoltaic cell~\cite{liu_high-performance_2016,chen_high-performance_2017,sharan_near-field_2025}. 

The purpose of the present work is not to replace full fluctuational-electrodynamics computations, but rather to derive a closed-form tunneling probability pertaining to near-field thermal exchange between dissimilar classes of materials that are found in practically relevant configurations that most often involve a metallic surface and an LED or PV cell. We derive a closed-form expression that explicitly demonstrates the role of material parameters, such as the plasma frequency of the metal, its damping constant, the background permittivity and the absorption strength of the semiconductor, as well as the vacuum gap size. We also derive the in-plane wave vector that dominates the radiative heat exchange and demonstrate that it is set jointly by the plasmonic and semiconductor media.

\section{Analytical model}

We consider two semi-infinite media separated by a vacuum gap of thickness $d$, with medium 1 representing a plasmonic material and medium 2 a semiconductor with a step-like absorption profile as outlined henceforth. For TPV applications, medium 2 is interpreted as the photovoltaic cell; for electroluminescent cooling it may represent the LED or photovoltaic element depending on the direction of operation, while for photodetection it represents the absorbing semiconductor interface. To maintain generality, no temperatures are assigned to either medium throughout the analysis. Consequently, the results do not rely on a prescribed thermal ordering between the semiconductor and plasmonic media, and are therefore applicable to TPV configurations, where the plasmonic medium is typically hotter, electroluminescent cooling configurations, where the semiconductor is cooled, as well as the rest of the aforementioned applications. In our formalism, the parallel component of the wave vector is normalized as $\beta=k_\parallel/k_0$, with $k_0=\omega/c$. For evanescent modes, $\beta>1$. The photon tunneling probability, as considered in the original Polder--van Hove formalism \cite{polder_theory_1971} can be written as 
\begin{equation}
\xi(\omega,\beta)=
\frac{4\,\mathrm{Im}\,[r_1]\,\mathrm{Im}\,[r_2]\, e^{-2\kappa_0 d}}
{\left|1-r_1 r_2e^{-2\kappa_0 d}\right|^2},
\qquad
\kappa_0=k_0\sqrt{\beta^2-1}.
\label{eq:xi_polder}
\end{equation}
Here, $r_1$ and $r_2$ are the Fresnel reflection coefficients of the two vacuum--material interfaces, defined for incidence from vacuum. Since this analysis focuses on evanescent surface-polaritonic coupling, which is predominantly transverse-magnetic (TM) polarized in local nonmagnetic media~\cite{maier_plasmonics_2007,joulain_surface_2005}, we retain only the TM contribution and denote the corresponding tunneling probability of the asymmetric emitter--absorber pair by $\xi_{\rm asy}$.

The spectral heat flux exchanged between the two media is obtained by 
\begin{equation}
\Phi_{\rm NF}(\omega)
=
\left[\Theta(\omega,T_{\rm H})-\Theta(\omega,T_{\rm C})\right]
\frac{k_0^2}{2\pi}
\Xi_{\rm exc}^{\rm NF}(\omega),
\label{eq:Phi_nf_spectral}
\end{equation}
where $\Theta(\omega,T)=\hbar\omega/[\exp(\hbar\omega/k_{\rm B}T)-1]$ is the mean energy per mode, and the temperatures $T_{\rm H}$ and $T_{\rm C}$ represent the equilibrium temperatures of the two materials and can be assigned interchangeably to either the plasmonic material or the semiconductor, depending on the application. The quantity $\Xi_{\rm exc}^{\rm NF}(\omega)$ is given by

\begin{equation}
\Xi_{\rm exc}^{\rm NF}(\omega)=
\int_0^{\infty}\beta\,\xi_{\rm asy}(\omega,\beta)\,d\beta.
\label{eq:Xi_nf}
\end{equation}
where $\xi_{\rm asy}(\omega,\beta)$ is the photon-tunneling probability of Eq.~\eqref{eq:xi_polder} for the considered asymmetric configuration. The quantity defining the modal and spectral distribution of the radiative heat transfer is thus the mode-resolved photon-tunneling probability $\xi_{\rm asy}$, which contains information about the dispersion characteristics of the plasmonic and semiconducting materials. 

We model the plasmonic material via the Drude model
\begin{equation}
\varepsilon_1(\omega)=\varepsilon_{\infty}
\left(1-\frac{\omega_{\rm p}^2}{\omega^2+i\gamma\omega}\right),
\label{eq:eps_drude}
\end{equation}
where $\omega_{\rm p}$ is the plasma frequency and $\gamma$ is the optical loss. The semiconducting material is modeled as a dielectric with a step-like onset of absorption at its bandgap, so that $\varepsilon_2(\omega)=\varepsilon_r$ for $\omega<\omega_g$ and $\varepsilon_2(\omega)=\varepsilon_r+i\varepsilon_{\rm i}$ for $\omega\geq\omega_g$. Although $\varepsilon_r$ and $\varepsilon_{\rm i}$ are treated as constants henceforth, the derivation applies to spectrally dispersive $\varepsilon_r$ and $\varepsilon_{\rm i}$ by replacing $\varepsilon_r$ and $\varepsilon_{\rm i}$ with $\mathrm{Re}[\varepsilon_2(\omega)]$ and $\mathrm{Im}[\varepsilon_2(\omega)]$. In the near field, TM-polarized evanescent modes dominate; hence, the reflection coefficients in Eq.~\eqref{eq:xi_polder} reduce to the quasistatic form $r_j^{\rm TM}\simeq(\varepsilon_j-1)/(\varepsilon_j+1)$. Substituting these expressions into Eq.~\eqref{eq:xi_polder}, with $\kappa_0\simeq\beta k_0$, yields
\begin{equation}
\boxed{
\xi_{\rm asy}(\omega,\beta)=
\frac{16\varepsilon_{\rm i}\gamma\varepsilon_{\infty}\omega\omega_{\rm p}^2}
{\Pi(\omega,\beta,d)}.
}
\label{eq:xi_closed}
\end{equation}

\begin{figure*}[t]
\includegraphics[width=\textwidth]{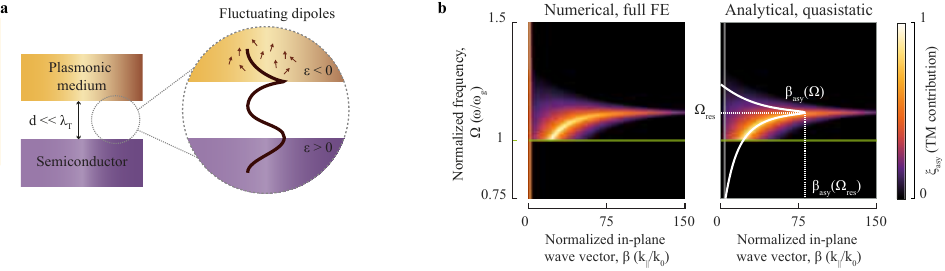}
\caption{
Analytical model and validation of photon tunneling in an asymmetric near-field heat-transfer system.
(a) Plasmonic-medium--vacuum-gap--semiconductor configuration considered in the model.
(b) TM photon-tunneling probability for an ITO--vacuum--InAs configuration, comparing the full fluctuational-electrodynamics calculation with the closed-form quasistatic expression in Eq.~\eqref{eq:xi_closed}. The predicted ridge $\beta_{\rm asy}(\omega)$ follows the dominant coupled surface-mode branch in the deep-evanescent regime. Panel (b) uses a reduced Drude damping $\gamma=5~{\rm meV}$ to make the coupled surface-mode branch visually clear.
}
\label{fig:model}
\end{figure*}

The denominator of $\xi_{\rm asy}(\omega,\beta)$ is given by
\begin{equation}
\Pi(\omega,\beta,d)=
e^{-2\beta k_0d}\mathcal A_-(\omega)
+\mathcal A_0(\omega)
+e^{2\beta k_0d}\mathcal A_+(\omega),
\label{eq:Pi_split}
\end{equation}
where
\begin{equation}
\begin{split}
\mathcal A_{\pm}(\omega)=&
\left[\varepsilon_{\rm i}^2+(\varepsilon_r\pm1)^2\right] \\
&\times
\left[\left((\varepsilon_{\infty}\pm1)\omega^2
-\varepsilon_{\infty}\omega_{\rm p}^2\right)^2
+\gamma^2(\varepsilon_{\infty}\pm1)^2\omega^2\right],
\end{split}
\label{eq:Apm}
\end{equation}
and 
\begin{equation}
\begin{split}
\mathcal A_0(\omega)=&
-2\left(-1+\varepsilon_{\rm i}^2+\varepsilon_r^2\right)\\
&\times
\left[(\varepsilon_{\infty}^2-1)\omega^4
-2\varepsilon_{\infty}^2\omega^2\omega_{\rm p}^2
+\varepsilon_{\infty}^2\omega_{\rm p}^4\right] \\
&+8\gamma\varepsilon_{\rm i}\varepsilon_{\infty}\omega\omega_{\rm p}^2.
\end{split}
\label{eq:Azero}
\end{equation}
Equations~\eqref{eq:xi_closed}--\eqref{eq:Azero} give the closed-form tunneling probability for the asymmetric cavity considered here, consisting of a plasmonic semi-infinite medium separated by a vacuum gap of thickness $d$ from a semi-infinitely thick semiconductor. The numerator of Eq.~\eqref{eq:xi_closed} is proportional to $\gamma\varepsilon_{\rm i}$, showing that both emitter loss and semiconductor absorption are required for heat exchange. As anticipated from the fluctuation-dissipation theorem, if either of the materials exhibits no loss, the photon-tunneling probability and related radiative heat exchange vanish. 

The three terms in $\Pi(\omega)$ (Eq.~\eqref{eq:Pi_split}) arise by expanding the denominator $\left|1-r_1r_2e^{-2\kappa_0d}\right|^2$ (Eq.~\eqref{eq:xi_polder}) by considering the Fabry--P\'erot-like multiple-reflections occurring within both media. The terms proportional to $e^{-2\beta k_0d}\mathcal A_-$ and $e^{2\beta k_0d}\mathcal A_+$ pertain to the two exponentially weighted material-response contributions, while $\mathcal A_0$ collects the interference term proportional to $\mathrm{Re}(r_1r_2)$. The dissipative contribution proportional to $\gamma\varepsilon_{\rm i}$ that appears both in the numerator of $\xi_{\rm asy}$ and in $\mathcal A_0$ links material absorption to the background permittivity ($\varepsilon_\infty$) and to the plasma frequency ($\omega_{\rm p}$) of the plasmonic material. 

To evaluate the closed-form expression, we consider a TPV system previously examined in the context of ideal near-field blackbody emitter--PV-cell coupling~\cite{papadakis_thermodynamics_2021}. The toy model in Ref.~\cite{papadakis_thermodynamics_2021} considered an indium tin oxide (ITO) plasmonic emitter coupled to an InAs PV cell, a material pair widely used to study plasmonic-emitter NF-TPVs~\cite{zhao_high-performance_2017,li_photon_2025}. Figure~\ref{fig:model}(b) compares the full fluctuational-electrodynamics calculation of the photon-tunneling probability, obtained using a transfer-matrix implementation following Ref.~\cite{legendre_crescent_2025}, with the analytical theory of Eqs.~\eqref{eq:xi_closed}--\eqref{eq:Azero} for a vacuum gap of $d=10~{\rm nm}$. The quasistatic approximation is valid when the relevant modes lie deep in the evanescent sector, $\beta\gg1$, and when the gap is subwavelength over the spectral range of interest, $k_0d\ll1$. These conditions are satisfied for the dominant branch shown in Fig.~\ref{fig:model}(b), and the analytical expression reproduces the full calculation in both frequency $\omega$ and normalized in-plane wave vector $\beta$. This comparison demonstrates that a real plasmonic--semiconductor cavity does not maintain $\xi=1$ over all useful frequencies and all wave vectors up to $1/(k_0d)$, as assumed in idealized near-field blackbody models. Instead, the enhancement is spectrally narrow, loss-dependent, and limited to a finite modal range~\cite{papadakis_thermodynamics_2021}.

\begin{figure*}[t]
\includegraphics[width=\textwidth]{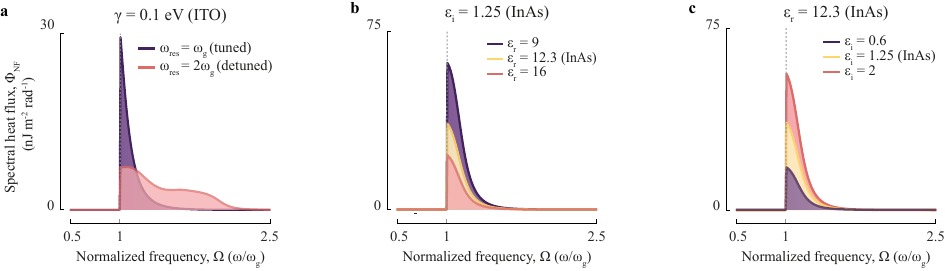}
\caption{
Material dependence of the spectral heat-flux density in an asymmetric near-field heat-transfer system.
(a) Effect of emitter tuning on the TM-polarized spectral heat-flux density $\Phi_{\rm NF}(\omega)$. When the coupled resonance is aligned with the semiconductor bandgap, $\omega_{\rm res}\approx\omega_g$, the exchanged spectrum is concentrated near the useful above-bandgap region; detuning shifts the exchanged power to higher photon energies.
(b) Dependence of $\Phi_{\rm NF}(\omega)$ on the real part of the semiconductor permittivity, $\varepsilon_r$, at fixed $\varepsilon_{\rm i}=1.25$.
(c) Dependence of $\Phi_{\rm NF}(\omega)$ on the imaginary part of the semiconductor permittivity, $\varepsilon_{\rm i}$, at fixed $\varepsilon_r=12.3$.
Unless otherwise stated, the calculations use $d=10~{\rm nm}$, $T_{\rm E}=1000~{\rm K}$, $T_{\rm C}=300~{\rm K}$, $\varepsilon_\infty=4$, $\omega_{\rm p}=0.45~{\rm eV}$, and an InAs-like reference semiconductor with $\varepsilon_r=12.3$ and $\varepsilon_{\rm i}=1.25$. Panels (a--c) use $\gamma=0.1~{\rm eV}$ unless otherwise indicated.
}
\label{fig:sweeps}
\end{figure*}

\section{Dominant in-plane wave vector and resonance condition}

A quantity that characterizes near-field radiative heat transfer is the resonant in-plane wave vector at which the photon-tunneling probability is maximized. Conventional estimates suggest that the largest contributing in-plane wave vector scales roughly as $1/d$, reflecting the exponential cutoff of evanescent modes across the vacuum gap~\cite{Pendry1999,Mulet2002}. However, this estimate is an approximation rather than a material-specific resonance condition, and it applies most naturally to symmetric cavities composed of low-loss plasmonic materials, such as coupled graphene sheets~\cite{Ilic2012}. It is therefore of merit to identify the resonant in-plane wave vector at which photon tunneling is maximized in the configuration considered here: an \textit{asymmetric} cavity.

From Eq.~\eqref{eq:xi_closed}, the numerator of $\xi_{\rm asy}(\omega,\beta)$ is independent of $\beta$; thus, at fixed $\omega$, the peak of $\xi_{\rm asy}(\omega,\beta)$ is controlled by the minimum of $\Pi$. Differentiating Eq.~\eqref{eq:Pi_split} yields 
\begin{equation}
e^{4\beta_{\rm asy}k_0d}=\frac{\mathcal A_-(\omega)}{\mathcal A_+(\omega)},
\end{equation}
such that
\begin{equation}
\beta_{\rm asy}(\omega)=
\frac{1}{4k_0d}
\ln\left[\frac{\mathcal A_-(\omega)}{\mathcal A_+(\omega)}\right].
\label{eq:beta_general}
\end{equation}
This stationary point corresponds to an evanescent mode only when it lies outside the light cone, in other words when $\beta_{\rm asy}>1$, or equivalently $\mathcal A_-/\mathcal A_+>e^{4\omega d/c}$.

Equation~\eqref{eq:beta_general} admits a simple interpretation in terms of two symmetric reference problems. Let $\beta_{\rm pl}(\omega)$ denote the dominant normalized in-plane wave vector obtained when both interfaces are described by the plasmonic response [Eq.~\eqref{eq:eps_drude}], and let $\beta_{\rm semi}(\omega)$ denote the corresponding quantity when both interfaces are described by the semiconductor response. The asymmetric plasmonic--semiconductor system then has a dominant normalized in-plane wave vector given by
\begin{equation}
\boxed{
\beta_{\rm asy}(\omega)=\frac{1}{2}\left[\beta_{\rm pl}(\omega)+\beta_{\rm semi}(\omega)\right].
}
\label{eq:beta_hybrid}
\end{equation}
The corresponding expressions for the two symmetric reference systems, $\beta_{\rm pl}$ and $\beta_{\rm semi}$, are provided in the Supplementary Material. Thus, the dominant in-plane wave vector of the asymmetric system can be interpreted as the average of two symmetric reference configurations: a plasmonic--plasmonic cavity and a semiconductor--semiconductor cavity. The coupled near-field mode is therefore not determined by the plasmonic medium alone, but by the joint electromagnetic response of the plasmonic and semiconductor media, and this is why it departs considerably from the $1/d$ estimate~\cite{Pendry1999,Mulet2002}.

The expression for $\beta_{\rm asy}$ identifies the normalized in-plane wave vector at which photon tunneling is maximized. This depends heavily on the plasmonic resonance, namely the frequency $\omega_{\rm sp}$ for which $\mathrm{Re}\,\varepsilon_1(\omega_{\rm sp})+1\simeq0$, where a plasmonic mode emerges at the interface with air. To identify the resonance frequency for photon tunneling between a semiconductor and a plasmonic medium, $\omega_{\rm res}$, within the quasistatic approximation, the condition changes to $\mathrm{Re}\,\varepsilon_1(\omega_{\rm res})+\varepsilon_r\simeq0$~\cite{zhao_high-performance_2017}, thereby redshifting the resonance frequency as long as $\varepsilon_r>1$. This expression underlies the surface-plasmon matching condition used in substrate-tailored near-field TPVs~\cite{nimje_critical_2025}, for which selecting the appropriate semiconductor and plasmonic emitter pair to satisfy this expression near the bandgap is more critical than maximizing overall near-field heat transfer. 

In Fig.~\ref{fig:sweeps}, we compute the spectral heat flux, Eq.~\eqref{eq:Phi_nf_spectral}, while varying the material parameters that enter the closed-form tunneling probability in Eq.~\eqref{eq:xi_closed}. These sweeps show that the analytical expression separates the roles of emitter tuning, semiconductor index, and semiconductor absorption in setting the spectral position, modal strength, and bandwidth of the near-field exchange. In Fig.~\ref{fig:sweeps}(a), varying the emitter plasma frequency shifts the coupled surface-mode resonance relative to the semiconductor bandgap: when $\omega_{\rm res}\approx\omega_g$, the heat flux is concentrated near the useful above-bandgap region, whereas detuning shifts it to higher photon energies. In Fig.~\ref{fig:sweeps}(b), varying $\varepsilon_r$ changes the semiconductor optical response without changing its absorption strength, allowing the role of refractive-index contrast to be isolated. Finally, Fig.~\ref{fig:sweeps}(c) shows that varying $\varepsilon_{\rm i}$ changes the semiconductor absorption entering the dissipative coupling required for heat exchange: increasing $\varepsilon_{\rm i}$ strengthens this coupling, but also broadens the resonance. Thus, the same material loss that enables photon tunneling also limits spectral selectivity.

The decomposition in Eqs.~\eqref{eq:xi_closed}--\eqref{eq:Pi_split} is obtained within the local, isotropic, quasistatic TM limit for semi-infinite media. More generally, the Polder--van Hove form in Eq.~\eqref{eq:xi_polder} applies to planar media described by reflection coefficients $r_1$ and $r_2$, which may depend on both frequency and in-plane wave vector. When these reflection coefficients reduce to scalar functions of frequency, the Fabry--P\'erot denominator $\left|1-r_1r_2e^{-2\kappa_0d}\right|^2$ can be rearranged into the denominator decomposition used here. The expression for the dominant in-plane wave vector in Eq.~\eqref{eq:beta_general}, however, requires the additional condition that $\mathcal A_-$, $\mathcal A_0$, and $\mathcal A_+$ do not depend explicitly on $\beta$. This condition is satisfied for local semi-infinite media, including Drude metals, Lorentz polar dielectrics, doped semiconductors, and the Drude-emitter--step-semiconductor model used in this work.

A useful consistency check is obtained in the symmetric plasmonic limit. If the semiconductor response in the analytical construction leading to Eq.~\eqref{eq:xi_closed} is replaced by the same plasmonic response as medium 1, such that $r_1=r_2=r$, the expression reduces to the evanescent tunneling probability used by Ref.~\cite{benabdallah_fundamental_2010} for two identical plasmonic media. Thus, the present analytical framework retains the same Fabry--P\'erot-like multiple-reflection denominator structure as the symmetric plasmonic case and extends it to a dissimilar plasmonic--semiconductor cavity, where $r_1\neq r_2$ and the dominant in-plane wave vector is set jointly by the two material responses. By contrast, the simple expression for the dominant in-plane wave vector generally fails for graphene, graphene-covered media, finite multilayers, and anisotropic or magneto-optical systems, where the reflection coefficients depend explicitly on $\beta$ or become matrices.

\section{Conclusion}

We derived a closed-form analytical expression for the photon-tunneling probability between a plasmonic interface and a semiconducting one. This asymmetric configuration is particularly relevant to near-field radiative heat transfer in several energy-conversion and optoelectronic applications, including near-field TPV systems~\cite{zhao_high-performance_2017,mittapally_near-field_2021,li_photon_2025}, electroluminescent cooling~\cite{liu_high-performance_2016,chen_high-performance_2017,sharan_near-field_2025}, and plasmon-assisted photodetection~\cite{sobhani_narrowband_2013}. We showed that the photon-tunneling probability is governed by a Fabry--P\'erot-like resonance condition that can be decomposed into three terms: two exponentially weighted terms that rely on materials' response, and an interference term [Eqs.~\eqref{eq:Pi_split}--\eqref{eq:Azero}]. We further showed that the dominant in-plane wave vector of tunneling can be interpreted as the average of the corresponding values for two symmetric reference systems, a plasmonic--plasmonic cavity and a semiconductor--semiconductor cavity [Eq.~\eqref{eq:beta_hybrid}]. This result reveals that near-field radiative exchange in the asymmetric cavity is not determined by either interface alone, but by the joint dielectric response of the plasmonic and semiconductor media. The derived model connects material parameters and vacuum gap size to the spectral position, linewidth, and modal extent of near-field radiative heat transfer. For TPV applications in particular, this emphasizes that enhanced heat transfer is useful only when evanescent modes are spectrally aligned with the photovoltaic bandgap and coupled to an absorbing semiconductor.

\section*{}
\begin{acknowledgments}
The authors thank Dr. Julien Legendre for constructive comments on the manuscript. This work was partially funded by CEX2024-001490-S (MICIU/AEI/10.13039/501100011033), the Spanish MICINN (PID2021-125441OA-I00, PID2020-112625GBI00, and CEX2019-000910-S), the European Union (Fellowship No. LCF/BQ/PI21/11830019 under Marie Sklodowska-Curie Grant Agreement No. 847648 and CATHERINA, 101168064), the Generalitat de Catalunya (2021 SGR 01443) through the CERCA program, Fundaci\'o Cellex, and Fundaci\'o Mir-Puig. Views and opinions expressed are those of the authors only and do not necessarily reflect those of the European Union or European Defence Agency.
\end{acknowledgments}

\bibliography{references}

\end{document}


\title{\textbf{Supplementary Material: \\ Analytical Theory of Photon Tunneling and Near-Field Heat Transfer Between Dissimilar Materials}}

\author{Kartika N. Nimje}
\affiliation{ICFO -- Institut de Ci\`encies Fot\`oniques, The Barcelona Institute of Science and Technology, Mediterranean Technology Park, Av. Carl Friedrich Gauss 3, Castelldefels, 08860, Barcelona, Spain}

\author{Mariano Pascale}
\affiliation{ICFO -- Institut de Ci\`encies Fot\`oniques, The Barcelona Institute of Science and Technology, Mediterranean Technology Park, Av. Carl Friedrich Gauss 3, Castelldefels, 08860, Barcelona, Spain}

\author{Georgia T. Papadakis}
\email{georgia.papadakis@icfo.eu}
\affiliation{ICFO -- Institut de Ci\`encies Fot\`oniques, The Barcelona Institute of Science and Technology, Mediterranean Technology Park, Av. Carl Friedrich Gauss 3, Castelldefels, 08860, Barcelona, Spain}

\date{\today}

\maketitle

\section{Algebraic origin of the denominator}
\label{sec:sm_derivation}

The three-term structure of the tunneling denominator follows from the Fabry--P\'erot-like multiple-reflection factor in the evanescent transmission probability,
\begin{equation}
\begin{split}
\left|1-r_1r_2e^{-2\kappa_0d}\right|^2
&=
1+|r_1r_2|^2e^{-4\kappa_0d} \\
&\quad -2e^{-2\kappa_0d}\mathrm{Re}(r_1r_2).
\end{split}
\label{eq:sm_fp_expansion}
\end{equation}
Using the quasistatic reflection coefficients $r_j^{\rm TM}\simeq(\varepsilon_j-1)/(\varepsilon_j+1)$, substituting the Drude emitter and step-absorbing semiconductor models, and multiplying by the common material denominator gives the form used in the main text,
\begin{equation}
\Pi(\omega,\beta,d)=
e^{-2\beta k_0d}\mathcal A_-(\omega)
+\mathcal A_0(\omega)
+e^{2\beta k_0d}\mathcal A_+(\omega).
\label{eq:sm_Pi_split}
\end{equation}
The two terms proportional to $\mathcal A_-$ and $\mathcal A_+$ originate from the exponentially weighted parts of Eq.~\eqref{eq:sm_fp_expansion}, while $\mathcal A_0$ originates from the interference term proportional to $\mathrm{Re}(r_1r_2)$.

\begin{figure}[t]
\includegraphics[width=\columnwidth]{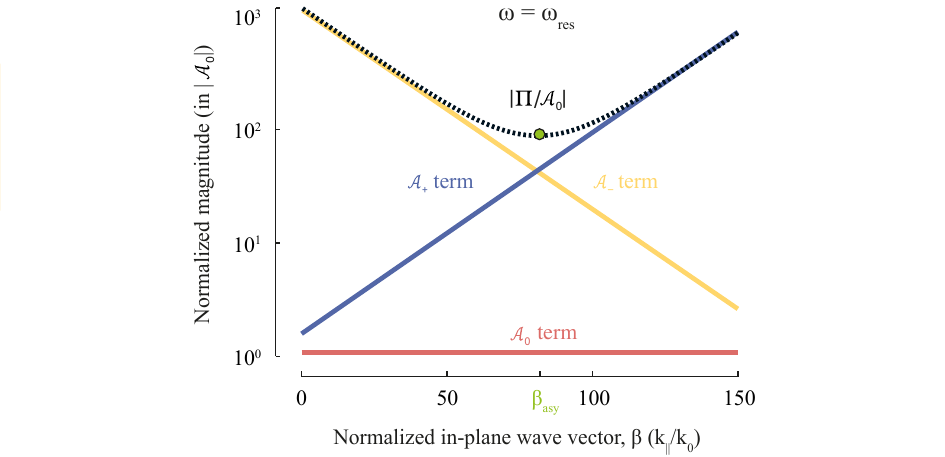}
\caption{
Decomposition of the analytical tunneling denominator at resonance. The two exponentially weighted contributions associated with $\mathcal A_-$ and $\mathcal A_+$, together with the interference contribution $\mathcal A_0$, produce a minimum near the dominant in-plane wave vector of the coupled surface mode.
}
\label{fig:sm_A_terms}
\end{figure}

Figure~\ref{fig:sm_A_terms} illustrates this decomposition at resonance. The two exponentially weighted terms produce a minimum in the denominator near the dominant in-plane wave vector, while $\mathcal A_0$ contains the interference contribution and the dissipative term associated with broadening.

\section{Expanded form of the dominant in-plane wave vector}
\label{sec:sm_beta}

Substituting the definitions of $\mathcal A_-$ and $\mathcal A_+$ into the stationary-point condition gives
\begin{widetext}
\begin{equation}
\beta_{\rm asy}(\omega)=
\frac{1}{4k_0d}
\ln\left\{
\frac{
\left[\varepsilon_{\rm i}^2+(\varepsilon_r-1)^2\right]
\left[
\left((\varepsilon_{\infty}-1)\omega^2-\varepsilon_{\infty}\omega_{\rm p}^2\right)^2
+\gamma^2(\varepsilon_{\infty}-1)^2\omega^2
\right]}
{
\left[\varepsilon_{\rm i}^2+(\varepsilon_r+1)^2\right]
\left[
\left((\varepsilon_{\infty}+1)\omega^2-\varepsilon_{\infty}\omega_{\rm p}^2\right)^2
+\gamma^2(\varepsilon_{\infty}+1)^2\omega^2
\right]}
\right\}.
\label{eq:sm_beta_asy_expanded}
\end{equation}
\end{widetext}

The two symmetric-reference wave vectors introduced in the main text are obtained by separating the plasmonic and semiconductor factors in Eq.~\eqref{eq:sm_beta_asy_expanded}. The plasmonic--plasmonic reference system gives
\begin{equation}
\begin{split}
\beta_{\rm pl}(\omega)
&=
\frac{1}{2k_0d}
\ln\left[
\frac{
\left((\varepsilon_{\infty}-1)\omega^2-\varepsilon_{\infty}\omega_{\rm p}^2\right)^2
+\gamma^2(\varepsilon_{\infty}-1)^2\omega^2}
{
\left((\varepsilon_{\infty}+1)\omega^2-\varepsilon_{\infty}\omega_{\rm p}^2\right)^2
+\gamma^2(\varepsilon_{\infty}+1)^2\omega^2}
\right],
\end{split}
\label{eq:sm_beta_pl}
\end{equation}
while the semiconductor--semiconductor reference system gives
\begin{equation}
\beta_{\rm semi}(\omega)
=
\frac{1}{2k_0d}
\ln\left[
\frac{(\varepsilon_r-1)^2+\varepsilon_{\rm i}^2}
{(\varepsilon_r+1)^2+\varepsilon_{\rm i}^2}
\right].
\label{eq:sm_beta_semi}
\end{equation}
These expressions satisfy
\begin{equation}
\beta_{\rm asy}(\omega)=
\frac{1}{2}\left[
\beta_{\rm pl}(\omega)+\beta_{\rm semi}(\omega)
\right].
\label{eq:sm_beta_hybrid}
\end{equation}

\section{Surface-mode resonance estimates}
\label{sec:sm_resonance_estimates}

The isolated plasmonic-emitter--vacuum surface mode satisfies
\begin{equation}
\mathrm{Re}[\varepsilon_1(\omega_{\rm sp})]+1\simeq0.
\label{eq:sm_sp_condition}
\end{equation}
For the Drude response, in the weak-loss limit, this gives
\begin{equation}
\omega_{\rm sp}
\simeq
\omega_{\rm p}
\sqrt{
\frac{\varepsilon_\infty}{\varepsilon_\infty+1}
}.
\label{eq:sm_omega_sp}
\end{equation}
If the same plasmonic response is referenced to a semiconductor environment of real permittivity $\varepsilon_r$, the corresponding quasistatic estimate becomes
\begin{equation}
\mathrm{Re}[\varepsilon_1(\omega_{\rm res})]+\varepsilon_r\simeq0,
\label{eq:sm_res_condition}
\end{equation}
or
\begin{equation}
\omega_{\rm res}
\simeq
\omega_{\rm p}
\sqrt{
\frac{\varepsilon_\infty}{\varepsilon_\infty+\varepsilon_r}
}.
\label{eq:sm_omega_res}
\end{equation}
For $\varepsilon_r>1$, the semiconductor-referenced resonance is redshifted relative to the isolated emitter--vacuum surface mode.

\section{Quasistatic gap scaling}
\label{sec:sm_scaling}

In the quasistatic short-gap regime, the relevant modes satisfy $\beta k_0d\lesssim1$, giving a characteristic cutoff $\beta_{\max}\sim1/(k_0d)$. If $\xi_{\rm asy}$ is approximately constant over this interval,
\begin{equation}
\Xi_{\rm exc}^{\rm NF}(\omega)
\sim
\int_1^{\beta_{\max}}\beta\,d\beta
\sim
\beta_{\max}^2
\sim
\frac{1}{(k_0d)^2}.
\end{equation}
The familiar $d^{-2}$ scaling therefore arises from the increasing number of high-$\beta$ modes available as the gap is reduced, rather than from an increase in the transmission of each individual mode. At sub-nanometer separations, this local-theory divergence is regularized by nonlocal response, surface roughness, and charge tunneling.

\bibliography{references}